\documentclass[twocolumn,showpacs,preprintnumbers,amsmath,amssymb,superscriptaddress,prb]{revtex4}

\usepackage{graphicx}
\usepackage{dcolumn}
\usepackage{bm}
\usepackage{color} 
\usepackage{ulem}

\begin{document}

\title{Strongly hybridized electronic structure of YbAl$_2$: An angle-resolved photoemission study}

\author{M.~Matsunami}
 \altaffiliation[Electronic address: ]{matunami@ims.ac.jp}
\affiliation{UVSOR Facility, Institute for Molecular Science, Okazaki 444-8585, Japan}
\affiliation{School of Physical Sciences, The Graduate University for Advanced Studies (SOKENDAI), Okazaki 444-8585, Japan}
\author{T.~Hajiri}
\author{H.~Miyazaki}
 \altaffiliation[Present address: ]{Center for Fostering Young and Innovative Researchers, Nagoya Institute of Technology, Nagoya 466-8555, Japan}
\affiliation{UVSOR Facility, Institute for Molecular Science, Okazaki 444-8585, Japan}
\author{M.~Kosaka}
\affiliation{Graduate School of Science and Engineering, Saitama University, Saitama 338-8570, Japan}
\author{S.~Kimura}
\affiliation{UVSOR Facility, Institute for Molecular Science, Okazaki 444-8585, Japan}
\affiliation{School of Physical Sciences, The Graduate University for Advanced Studies (SOKENDAI), Okazaki 444-8585, Japan}

\date{\today}

\begin{abstract} 
We report the electronic structure of a prototypical valence fluctuation system, YbAl$_2$, using angle-resolved photoemission spectroscopy. 
The observed band dispersions and Fermi surfaces are well described in terms of band structure calculations based on local density approximation. 
Strong hybridization between the conduction and 4$f$ bands is identified on the basis of the periodic Anderson model. 
The evaluated small mass enhancement factor and the high Kondo temperature qualitatively agree with those obtained from thermodynamic measurements. 
Such findings suggest that the strong hybridization suppresses band renormalization and is responsible for the valence fluctuations in YbAl$_2$. 
\end{abstract}


\maketitle

Heavy-fermion or valence fluctuation systems are characterized by the strong correlations of $f$ electrons, resulting in an enhanced effective mass of quasiparticles due to the band renormalization effect \cite{Hewson}. 
Consequently, the band structure and Fermi surface (FS) topology are modified from those predicted in band structure calculations based on local density approximation (LDA). 
To estimate the band renormalization effect, a direct comparison of the experimental band structure as well as the FS with the calculations is necessary. 
Angle-resolved photoemission spectroscopy (ARPES), which can probe both band structure and FS, is the most suitable tool for this purpose.

Another important concept in heavy-fermion systems is a hybridization effect between conduction ($c$) and $f$ electrons. 
The $c$-$f$ hybridization-derived electronic structure has been verified by ARPES studies \cite{Arko,Danzenbacher,Wigger,Vyalikh,Mo,Im,Koitzsch,Klein,Denlinger,Yano,Okane,Yasui}. 
However, most of those studies have been performed on the basis of the surface Brillouin zone (BZ) \cite{Arko,Danzenbacher,Wigger,Vyalikh,Mo} or by using the 4$d$-4$f$ resonant process \cite{Im,Koitzsch,Klein}, in which $k_z$ (the momentum normal to surface) dependence on the electronic structure is neglected owing to its quasi-two-dimensionality. 
In general, ARPES along the bulk BZ with well-defined $k_z$ should be ideal for the three-dimensional materials. 
Furthermore, ARPES data acquired at the high-symmetry point in the bulk BZ can be closely compared with the band structure calculations. 
In the case of Ce-based compounds, however, resonant photoemission is intrinsically required to directly probe the $f$-derived feature in the bonding band of the $c$-$f$ hybridized bands because of the small $f$-electron occupation number ($n_f$$\leq$1). 
On the other hand, in Yb-based compounds, the large occupation number ($n_f$$\geq$13) enables us to probe the dispersions of not only the bonding band but also the antibonding band without any resonant process, and hence the photon-energy ($h\nu$) tuning of $k_z$ into the high-symmetry point is available. 
The Kondo resonance peak also exists on the occupied side, in contrast to Ce compounds and hence is accessible with ARPES.

Here, we focus on YbAl$_2$, which is known as one of the prototypical valence fluctuation systems. 
In a recent hard x-ray photoemission spectroscopy study, the mean valence of Yb ions was estimated as +2.2 below 300\,K \cite{HAXPES}. 
Correspondingly, an extremely high Kondo temperature ($T_{\rm K}$) exceeding 2000\,K has been suggested by the magnetic susceptibility \cite{Suscep} or the inelastic neutron scattering \cite{neutron}. 
The mass enhancement factor is particularly small among heavy-fermion systems known so far \cite{Tsujii-KW}, implying the small renormalization effect. 
Thus, YbAl$_2$ can be a suitable system to investigate the applicable limit of LDA calculation in relation with the fluctuating valence for the heavy-fermion systems.

In this paper, we report the results of ARPES performed on YbAl$_2$. 
The band structure and FSs obtained are comparable to those of the LDA calculation. 
The $c$-$f$ hybridized band dispersions near the Fermi level ($E_{\rm F}$) are clearly observed and analyzed on the basis of a periodic Anderson model (PAM) without interaction. 
The estimated mass enhancement factor and $T_{\rm K}$ are consistent with the other experiments. 
These results suggest that the strong $c$-$f$ hybridization suppresses the band renormalization and hence the electron correlation effect of Yb\,$4f$ states does not work significantly in YbAl$_2$, even though rare-earth compounds are generally believed to have a strong correlation effect.

Single crystals of YbAl$_2$ were grown by the lithium flux method \cite{Nowatari}. 
The ARPES experiment was carried out at the undulator beamline BL7U of UVSOR-II in the Institute for Molecular Science, with $h\nu$ = 16--29\,eV and using a hemispherical electron analyzer (MBS A-1, MB Scientific) \cite{SAMRAI}. 
The total energy resolution and the measurement temperature were set to 20\,meV and 12\,K, respectively. 
The vacuum was maintained below 5$\times$10$^{-9}$\,Pa during the experiment. 
The crystal orientation was determined by back-reflection Laue x-ray diffraction prior to the ARPES measurements. 
Clean sample surfaces for ARPES measurements were obtained by cleaving $in$ $situ$ at 12\,K along the (111) plane. 
The LDA calculation was performed by the full-potential linearized-augmented-plane-wave plus local-orbital (LAPW+lo) method including spin-orbit coupling implemented in the WIEN2k code \cite{WIEN2k}.

\begin{figure}[t]
\begin{center}
\includegraphics[width=0.46\textwidth]{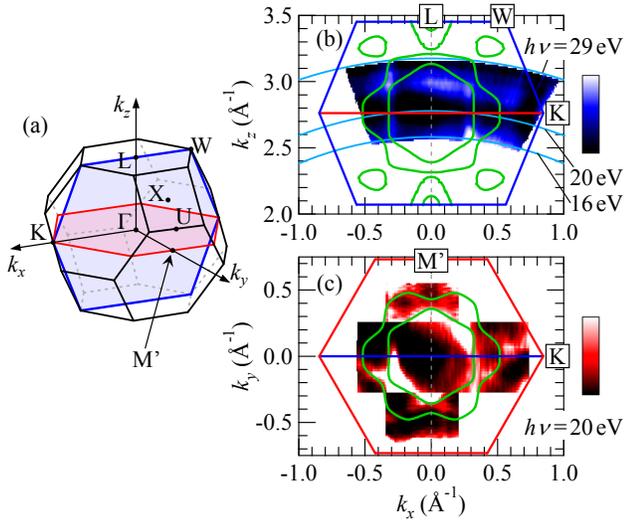}
\caption{
(Color online) 
(a) BZ of YbAl$_2$, in which $k_z$ corresponds to the (111) direction of the fcc BZ. 
In the $k_x$-$k_y$ plane, the midpoint of the zone edge ($K$) is referred to as $M'$ point, which is also equivalent to the midpoint between $L$ and $U$ points. 
(b) and (c) FS mapping in the $k_x$-$k_z$ and $k_x$-$k_y$ planes, respectively. 
The FS is obtained by a plot of the integrated intensity within $E_{\rm F}$$\pm$50\,meV in the ARPES spectra. 
The color scale is shown at the right of each map. 
The FSs obtained by the LDA calculation are overlapped in (b) and (c). 
} 
\end{center}
\end{figure}

Figure~1(a) shows the BZ of YbAl$_2$, in which $k_z$ corresponds to the (111) direction of fcc BZ. 
The FS in the $k_x$-$k_z$ plane is obtained by the $h\nu$-dependent ARPES, as shown in Fig.~1(b). 
The inner potential is determined to be 13.7\,eV from the symmetry of the FS-image contrast. 
Figure~1(c) shows the ``in-plane'' FS in the $k_x$-$k_y$ plane, measured at $h\nu$ = 20\,eV. 
The FSs obtained by the LDA calculation are also plotted in Figs.~1(b) and 1(c). 
Both inner and outer FSs around the $\Gamma$ point, as predicted in the LDA calculation, are observed in both the $k_x$-$k_z$ and $k_x$-$k_y$ planes. 
A further comparison between the experimental results and calculations is given below on the basis of the band structure.

Figures 2(a) and 2(b) show an intensity plot of the ARPES spectra taken along the $\Gamma$-$K$ direction and their angle-integrated spectrum, respectively, measured at $h\nu$ = 20\,eV. 
Both the widely dispersive $c$ bands and the nearly dispersionless $f$ bands are clearly visible in the ARPES image. 
Regarding the $f$ bands, which are of the Yb$^{2+}$ component corresponding to the 4$f^{13}$ final state, the main features at the binding energy ($E_{\rm B}$) of 0.17 and 1.49\,eV are attributed to the spin-orbit split components 4$f_{7/2}$ and 4$f_{5/2}$, respectively, in the bulk. 
The 4$f_{7/2}$ peak was fitted by a Lorentzian multiplied by a Fermi-Dirac distribution (FDD) and an integral background, as shown by the thin solid line in Fig.~2(b). 
The peak position and width obtained were 0.171 and 0.155\,eV, respectively, which are consistent with those in the hard and soft x-ray photoemission spectra \cite{HAXPES}. 
Additional fine structures can be seen at $E_{\rm B}$$\sim$1.09\,eV and 1.25\,eV, which might be derived from the subsurface and surface 4$f_{7/2}$ states, respectively \cite{PES_Kaindl}. 
Such non-bulk components have little impact on the bulk 4$f_{7/2}$ band, which is discussed in more detail in the following.

\begin{figure}[t]
\begin{center}
\includegraphics[width=0.46\textwidth]{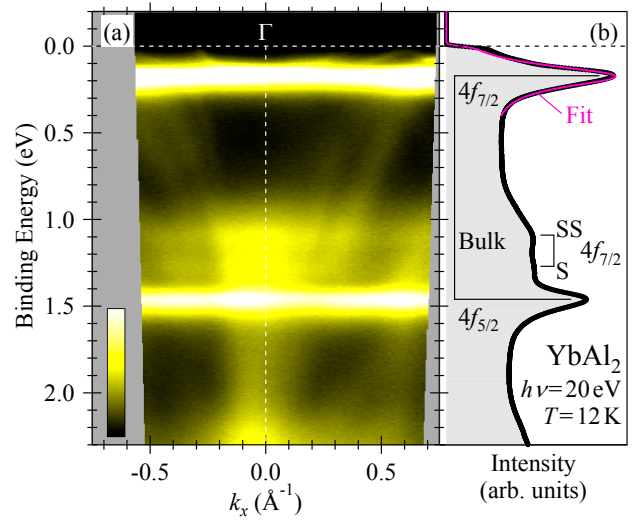}
\caption{
(Color online) 
(a) Intensity plot of the ARPES spectra of YbAl$_2$ taken along the $\Gamma$-$K$ direction. 
The color scale is shown at the bottom left. 
(b) Angle-integrated spectrum obtained from (a). 
The 4$f_{7/2}$ peak is fitted by a Lorentzian multiplied by an FDD and an integral background. 
``S'' and ``SS'' denote the surface and subsurface states, respectively. 
} 
\end{center}
\end{figure}

\begin{figure}[t]
\begin{center}
\includegraphics[width=0.46\textwidth]{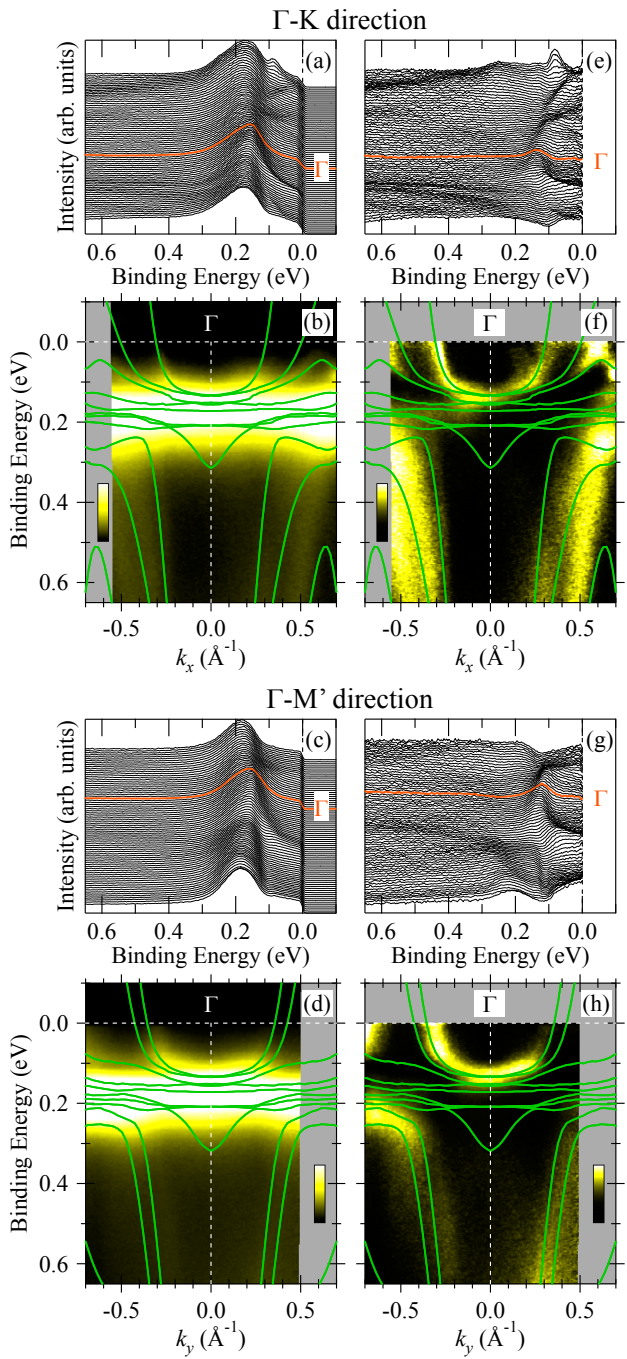}
\caption{
(color online) 
(a) and (c) Series of ARPES spectra near $E_{\rm F}$ along the $\Gamma$-$K$ and $\Gamma$-$M'$ directions, respectively. 
(b) and (d) ARPES intensity plots of (a) and (c), respectively. 
(e)--(h) The same data as in (a)--(d), normalized to the angle-integrated spectrum. 
The band dispersions obtained by the LDA calculation are overlapped in (b), (d), (f), and (h). 
} 
\end{center}
\end{figure}

Figures 3(a)-3(d) show the series of ARPES spectra and the ARPES intensity plot near $E_{\rm F}$ along the $\Gamma$-$K$ direction and the direction from $\Gamma$ to the midpoint of the zone edge ($K$), hereafter, referred to as $M'$ point. 
The band dispersions obtained by the LDA calculation are also plotted in Figs.~3(b) and 3(d). 
The observed Yb\,4$f_{7/2}$ band around $E_{\rm B}$ = 0.17\,eV shows good agreement of energy with the calculation without any shift. 
The degree of coincidence is much greater than that of the other nearly divalent Yb system YbCu$_2$Ge$_2$ \cite{Yasui}. 
To emphasize the $c$-band dispersions, each energy distribution curve (EDC) in the ARPES image was normalized to the angle-integrated spectrum. 
Alternatively, the same results can be achieved by normalization for each momentum distribution curve (MDC) area. 
The ARPES spectra and image obtained, shown in Figs.~3(e)--3(h), can be regarded as ``effective off-resonance'' ARPES spectra and image, representing hybridized-band dispersions. 
It should be noted that, unlike purely divalent systems \cite{Yb2+_Matsunami}, the existence of these hybridized bands is responsible for the valence fluctuations in YbAl$_2$. 
Their overall feature is well described in terms of the LDA calculation. 
For the $\Gamma$-$K$ direction, among the anti-bonding bands located in the lower $E_{\rm B}$ region compared with the 4$f_{7/2}$ band, the innermost band matches very well. 
On the other hand, the experimental outer band has a slightly weaker dispersion. 
This band might not be separated from the more outer band, which does not cross $E_{\rm F}$ as shown in the LDA calculation. 
A similar slight deviation from the LDA calculation is recognized in the bonding bands shown in Fig.~3(f). 
For the $\Gamma$-$M'$ direction, the experimental inner and outer bands are highly mixed. 
The ``mixed'' band is located on the inside of the calculated one and has almost the same Fermi vector ($k_{\rm F}$) as the inner band in the $\Gamma$-$K$ direction. 
This finding suggests that the inner FS is more circular than that of the LDA calculation, as expected from Fig.~1(c).

\begin{figure}[t]
\begin{center}
\includegraphics[width=0.46\textwidth]{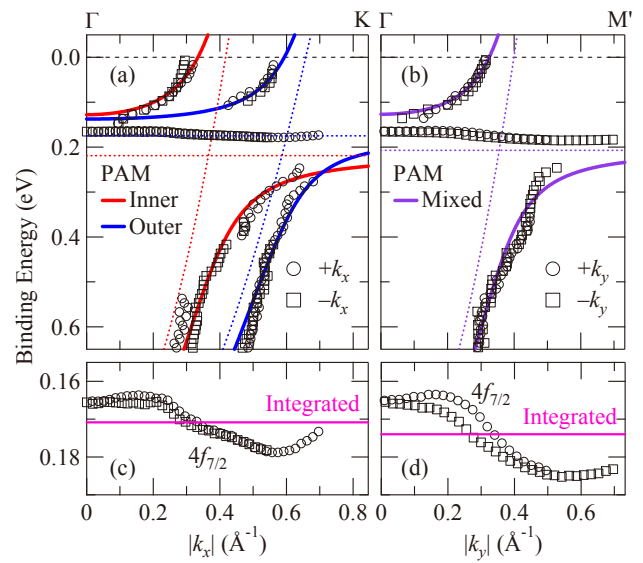}
\caption{
(Color online) 
(a) and (b) Analysis of band dispersions along the $\Gamma$-$K$ and $\Gamma$-$M'$ directions, respectively, by $c$-$f$ hybridized bands based on PAM. 
The dotted lines indicate the $c$ bands and $f$ states before hybridization. 
The experimental data points were obtained by a Lorentzian fitting of MDC or EDC. 
(c) and (d) Expansion of dispersion for the 4$f_{7/2}$-band. 
The angle-integrated data were obtained by the fitting of the angle-integrated spectra.
} 
\end{center}
\end{figure}

In order to quantify the impact of the $c$-$f$ hybridization effect apart from the LDA calculation, the observed band dispersions were analyzed in terms of PAM. 
In the non-interacting case with $U_{ff} = 0$, the band dispersion of PAM is given by the form of hybridized bands as 
\begin{equation}
E_k^{\pm}=\frac{\varepsilon_k+E_{f, k}\pm\sqrt{\mathstrut (\varepsilon_k-E_{f, k})^2+4V_k^2}}{2}, 
\end{equation}
where $\varepsilon_k$, $E_{f, k}$, and $V_k$ are the $c$-band, $f$-band, and hybridization energies, respectively \cite{Hewson}. 
As a simplification to fit the experimental data using this formula, a parabolic electron band was assumed for $\varepsilon_k$ and the $k$ dependence of $E_{f, k}$ and $V_k$ was neglected. 
The fitting results are shown in Figs.~4(a) and 4(b), in which the experimental data points were obtained by a Lorentzian fitting of MDC or EDC and are plotted as the absolute value of $k_x$ or $k_y$. 
The band dispersions are well reproduced by this simple model with the parameters summarized in Table~I. 
From the value of $E_{f}$, $T_{\rm K}$ is evaluated as 2540, 2030, and 2400\,K for the inner, outer, and mixed bands, respectively. 
In this case, the lowest $T_{\rm K}$ is meaningful and consistent with that obtained by other experiments \cite{Suscep,neutron}. 
The mass enhancement is simply related to the dispersion of $E_k^+$ and $\varepsilon_k$ at each $k_{\rm F}$ by the relation 
\begin{equation}
\frac{m^*}{m_b}=\frac{\partial^2 \varepsilon_k}{\partial k^2} \Big/ \frac{\partial^2 E_k^+}{\partial k^2}, 
\end{equation}
where $m_b$ is the unhybridized band mass. 
For each band, the mass enhancement factor $m^*/m_b$ is evaluated as listed in Table.~I. 
Since the electronic specific heat coefficient $\gamma$ of YbAl$_2$ has been reported as $\sim$10--17\,mJ/K$^2$mol \cite{Nowatari,Fisk}, and that of LuAl$_2$ as a reference is 5.7\,mJ/K$^2$mol \cite{LuAl2_Gamma}, the ratio, $\gamma$(YbAl$_2$)/$\gamma$(LuAl$_2$) = 1.8--3, is comparable to the mass enhancement obtained in our ARPES study ($m^*/m_b$ = 1.42--2.79). 
It is quite reasonable that the outer band, with weak dispersion compared with the LDA calculation, has a heavier mass and a weaker hybridization.

\begin{table}[tbp]
\caption{
Parameters $E_f$ and $V$ used to fit the band dispersions in Figs.~4(a) and 4(b) via Eq.~(1), and the evaluated mass enhancement factor $m^*/m_b$. 
}
\begin{ruledtabular}
\begin{tabular}{lllll}
Direction & Band & $E_f$\,(eV) & $V$\,(eV) & $m^*/m_b$ \\ 
\hline
$\Gamma$-$K$ & Inner & 0.219 & 0.272 & 1.42 \\
$\Gamma$-$K$ & Outer & 0.175 & 0.186 & 2.79 \\
$\Gamma$-$M'$ & Mixed & 0.207 & 0.263 & 1.51 \\
\end{tabular}
\end{ruledtabular}
\end{table}

The mass enhancement observed in YbAl$_2$ can be described in terms of a $c$-$f$ hybridization effect. 
On the other hand, the renormalization effect due to the electron correlation $U_{ff}$ cannot contribute significantly. 
Even if the parameters obtained by this PAM analysis include a renormalization effect in some manner such as $\tilde E_f$ and $\tilde V$, the impact would be very small, as expected from the degree of mass enhancement. 
More importantly, the coincidence of the $4f_{7/2}$ band position with the LDA calculation strongly suggests an extremely small renormalization. 
It is also consistent with the small renormalization factor suggested in the optical conductivity of YbAl$_2$ \cite{Kimura}. 
In the overall valence band, the localized $f$ part at $E_{\rm B}$ = 5--12\,eV corresponding to the $4f^{12}$ final state, which is separated by $U_{ff}$ from the unoccupied $4f^{14}$ final state, has a very weak intensity as evidenced by the Yb valence \cite{HAXPES,PES_Kaindl}. 
Therefore, PAM analysis under the condition of $U_{ff} = 0$ can be useful for YbAl$_2$.

Finally, we consider the 4$f_{7/2}$ peak in connection with a Kondo resonance peak, not merely a shallow core level. 
As shown in Figs.~4(c) and 4(d), this peak shows a small but clear dispersion \cite{4f_7/2}. 
Such behavior, which is also observed for the strongly correlated YbRh$_2$Si$_2$ \cite{Wigger,Vyalikh,Mo}, is consistent with the theoretical predictions on the basis of PAM \cite{PAM-PES}. 
As a possible point of view, therefore, this finding can be regarded as an evidence for the dispersion of a resonance peak in the coherent Kondo ground state of YbAl$_2$. 
On the other hand, theoretically, the ``Kondo'' resonance is defined only in the ``Kondo lattice'' regime (near Yb$^{3+}$ state), not in the valence fluctuation (mixed valence) regime such as YbAl$_2$ \cite{Kumar}. 
The relevant issues have been discussed on the basis of several photoemission studies of YbAl$_3$, which has larger Yb valence than that of YbAl$_2$ \cite{YbAl3-1,YbAl3-2}. 
Indeed, the identification of the Kondo effect, which represents a typical many-body phenomenon \cite{Hewson}, may seem to be unreasonable for YbAl$_2$ exhibiting an extremely small renormalization effect. 
In Yb-based compounds, however, the valence fluctuation state with the energy scale of $T_{\rm K}$ can appear over a broad range of Yb valence between $+$2 and $+$3, based on the fact that the spin-orbit splitting of Yb\,4$f$ electrons is generally larger than the 4$f$ bandwidth \cite{Flouquet}. 
This contrasts with the Ce case, in which the valence fluctuation is restricted near $+$3 since the Ce\,4$f$ spin-orbit splitting is smaller than the 4$f$ bandwidth. 
In the case of YbAl$_2$, the Yb valence is nearly divalent $+$2.2 as a consequence of the strong $c$-$f$ hybridization \cite{HAXPES}. 
With decreasing hybridization strength toward Yb$^{3+}$, the resonance peak will move closer to $E_{\rm F}$ and become sharper, providing a well-defined Kondo resonance peak \cite{Kumar,Costi}. 
Such an evolution of Kondo resonance may bring about the drastically modified FS topology and the highly renormalized band structure compared with the LDA calculation. 
In contrast, at the Yb$^{2+}$ limit, the resonance peak turns into the shallow core level of the 4$f_{7/2}$ states and does not contribute to the FS \cite{Yb2+_Matsunami}. 
As a result, for YbAl$_2$, the coherent Kondo state at much lower temperature than $T_{\rm K}$ coexists sufficiently well with the nearly unrenormalized band structure to match with the LDA calculation.

In conclusion, we have performed ARPES of a prototypical valence fluctuation system YbAl$_2$. 
The band dispersions obtained around the $\Gamma$ point, particularly for the energy position of the Yb\,4$f_{7/2}$ band, are well described by the LDA calculation without any modification. 
This may indicate that the electron correlation of Yb\,$4f$ states does not work significantly. 
A strongly $c$-$f$ hybridized nature is identified on the basis of noninteracting PAM. 
These results suggest a well-balanced coexistence between the coherent Kondo state and the nearly unrenormalized band structure in YbAl$_2$.

We would like to thank T. Ito for fruitful discussions. 
This work was performed by the Use-of-UVSOR Facility Program of the Institute for Molecular Science (2011).


\begin{thebibliography}{99}
\bibitem{4f_7/2}
The 4$f_{7/2}$ state has eightfold degeneracy. 
Our findings indicate that the two or three orbitals among them are involved in the hybridization and the residual orbitals remain nearly unhybridized and dispersionless.
\end{thebibliography}

\end{document}